\documentclass[aps,prl,groupedaddress,reprint,superscriptaddress]{revtex4-1}

\usepackage{amsfonts}
\usepackage{graphicx}
\usepackage{mathcomp}
\usepackage{wasysym}
\newcommand{\etal}{\em et al. \rm}

\begin{document}
\title{Long-lived, radiation-suppressed superconducting quantum bit in a planar geometry }

\author{Martin Sandberg}

\email[]{Electronic addreass: martin.sandberg@nist.gov}
\affiliation{National Institute of Standards and Technology, Boulder, Colorado 80305, USA}

\author{Michael R. Vissers}

\affiliation{National Institute of Standards and Technology, Boulder, Colorado 80305, USA}

\author{Tom Ohki}


\affiliation{Raytheon BBN Technologies, Cambridge, Massachusetts 02138, USA}

\author{Jiansong Gao}
\affiliation{National Institute of Standards and Technology, Boulder, Colorado 80305, USA}

\author{Jose Aumentado}
\affiliation{National Institute of Standards and Technology, Boulder, Colorado 80305, USA}

\author{Martin Weides}
\altaffiliation{Current address: Karlsruhe Institute of Technology, 76131 Karlsruhe, Germany}
\affiliation{National Institute of Standards and Technology, Boulder, Colorado 80305, USA}


\author{David P. Pappas}

\email[]{Electronic address: david.pappas@nist.gov}
\affiliation{National Institute of Standards and Technology, Boulder, Colorado 80305, USA}

\begin{abstract}
We present a superconducting qubit design that is fabricated in a 2D geometry over a superconducting ground plane to enhance the lifetime. The qubit is coupled to a microstrip resonator for readout. The circuit is fabricated on a silicon substrate using low loss, stoichiometric titanium nitride for capacitor pads and small, shadow-evaporated aluminum/aluminum-oxide junctions. We observe qubit relaxation and coherence times ($T_1$ and $T_2$) of 11.7 $\pm$ 0.2 $\mu$s and 8.7 $\pm$ 0.3 $\mu$s, respectively. Calculations show that the proximity of the superconducting plane suppresses the otherwise high radiation loss of the qubit. A significant increase in $T_1$ is projected for a reduced qubit-to-superconducting plane separation.
\end{abstract}

\maketitle 
Superconducting qubits are a leading candidate for implementing scalable quantum information processing. A wide range of important experiments, such as violation of Bell's inequality \cite{Ansmann2009,Palacios-Laloy2010,Mariantoni2011}, two qubit algorithms \cite{Mariantoni2011,Steffen2006,DiCarlo2009}, and implementation of error correction codes \cite{Reed2012,Fedorov2012} have been demonstrated using these circuits. Since the first experiment by Nakamura \etal \cite{Nakamura1999}, significant effort has been expended to improve relaxation and coherence times, i.e. $T_1$ and $T_2$. For example, optimal biasing increased $T_2$ times \cite{Vion2002} for charge qubits, and further improvements in both $T_1$ and $T_2$ were obtained using a device design that is inherently insensitive to charge noise, known as the transmon qubit \cite{Schreier2008,Koch2007}. In these devices the readout is integral to the operation of the qubit. One widely used scheme for readout is the circuit quantum electrodynamics (cQED) architecture \cite{Wallraff2004}, where the qubit is dispersively coupled to a microwave resonator. In addition to its function as a measurement element, the resonator also acts as a filter of the electromagnetic environment. This reduces the number of decay channels for the qubit, thereby improving the qubit lifetime \cite{Blais2004}. Typical lifetimes for a transmon qubit coupled to a 2D co-planar waveguide (CPW) resonator are on the order of a few $\mu$s, although lifetimes as long as 9.7 $\mu$s have been demonstrated by going to large area interdigitated capacitor plates \cite{Chow2012}, in order to reduce material loss.

Recent experiments have demonstrated considerably longer lifetimes for transmon qubits embedded in 3D high quality-factor cavities \cite{Paik2011}. The cavity serves to both suppress radiation from the qubit and measure the qubit state. Using these cavity resonators, $T_1$ and $T_2$ times approaching 100 $\mu$s have been reported \cite{Rigetti2012}. As pointed out in Ref. \cite{Paik2011}, these long lifetimes show that the Josephson junction is not the primary source of energy decay observed in the planar circuits. This indicates that, with proper engineering, planar Josephson circuits with long lifetimes should also be realizable.  
\begin{figure}[tb]
\includegraphics[width=8.6cm]{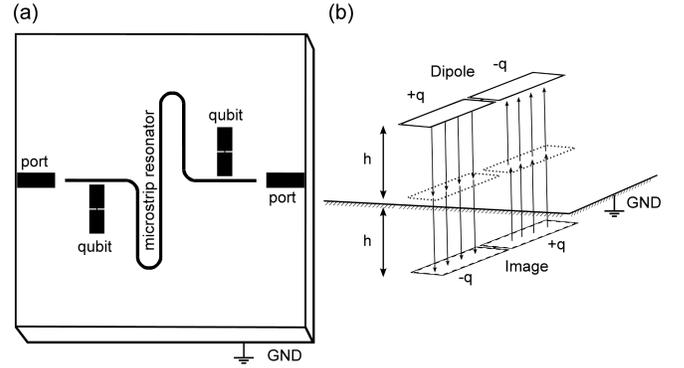}
\caption{\label{Chip_fig}(a) Chip layout. The device consists of two transmon qubits capacitively coupled to a microstrip resonator with a continuous superconducting ground plane on the backside. The resonator and the large qubit pads are made from TiN whereas the Josephson junction interconnecting the pads is made from Al/AlO$_x$/Al (b) Illustration of a dipole (the qubit in this case) at the distance h above a superconducting plane. The plane generates a mirror image of opposite charge ($\pm$q) which acts to suppress the radiation from the dipole.}
\end{figure}

In this work, we present an improved, 2D planar geometry for a qubit in a cQED architecture. Our design uses a microstrip resonator to read out the qubit, as shown in FIG. \ref{Chip_fig}(a). This geometry has several benefits. Most importantly, the addition of a superconducting plane on the back side of the chip suppresses radiative loss from the qubit. It also  allows for elimination of discontinuous ground planes on the top side of the chip that can cause stray resonances.

The qubits are of the transmon type, i.e. the ratio of Josephson energy ($E_J$) to the charging energy ($E_c$) $\sim100$. The design used here is similar to that used by Paik \etal \cite{Paik2011}. The circuits are fabricated primarily from stoichiometric titanium nitride (TiN) on intrinsic silicon (Si). Titanium nitride on Si is used because of its low microwave loss; CPW resonators made from TiN on Si have internal quality factors greater than $1\times 10^6$ at single photon excitation \cite{Vissers2010}. The very low loss makes the TiN-Si system ideal for quantum circuits.
 
The TiN was deposited consecutively onto the top and bottom surfaces of a hydrogen terminated Si wafer (at 500 $^{\circ}$C) using reactive sputter deposition  \citep{Vissers2010}. The capacitor plates and the microstrip resonator were patterned into the top film with the use of photolithography. The structures were then fabricated in three steps. First, a small area was opened up where the junction was to be placed. This was done using a highly controllable CF$_4$-based reactive ion etch (RIE). In the second step, the remaining TiN circuit was etched using a SF$_6$-based RIE. The second step is necessary because, while the SF$_6$ etch produces low loss Si surfaces \cite{Sandberg2012}, it also produces large trenches (due to a high etch rate, 20:1, of Si:TiN in SF$_6$) that are not suitable for the junction area. In the third step, the Josephson junction interconnect between the capacitor plates was patterned with electron-beam lithography and formed by use of double angle Al evaporation and oxidation.

The sample was mounted in a Cu sample box with the TiN backplane electrically connected to the box. The ports (see Fig. \ref{Chip_fig} (a)), were wire-bonded to circuit boards leading to the coaxial connectors of the sample box.      
The sample was measured at the base temperature of a $^3$He/$^4$He dilution refrigerator ($\approx$ 20 mK). 
To characterize the two qubits (q1 and q2) on the chip, we used a pulsed, strong readout tone, resonant with the bare resonator frequency of 6.55 GHz \cite{Reed2010}. The eigen-frequencies of the qubits were found to be $\nu_{q1}=$ 5.24 GHz and $\nu_{q2}=$ 6.18 GHz. We extracted a charging energy $E_c/h\approx 270$ MHz from the two photon Rabi oscillations of the 0$\rightarrow$2 qubit transition and a coupling strength of $g/h\approx 150 $ MHz from the dispersive shift of the cavity. From these, we obtain a Josephson energy of $E_J/h \approx 15$ GHz and $E_J/h\approx 21$ GHz for both q1 and q2. 

\begin{figure}[tb]
\includegraphics[width=8.6cm]{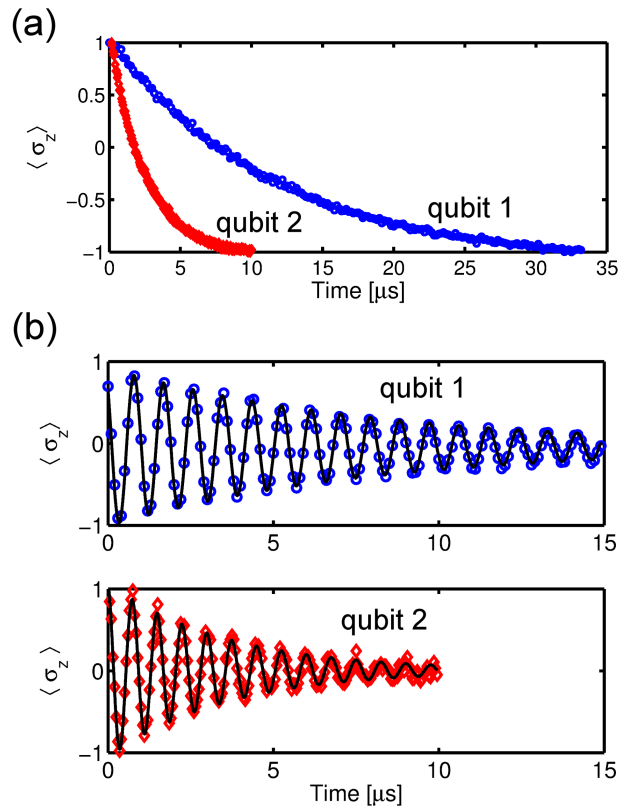}
\caption{\label{Time_fig}(a) Measured relaxation time for the two qubits. We extract a $T_1$ time of 11.7 $\pm$ 0.2 $\mu$s and 2.1 $\pm$ 0.1 $\mu$s  for qubit 1 and qubit 2, respectively. The $T_1$ time for qubit 2 agrees well with the expected Purcell limit, whereas $T_1$ for qubit 1 is shorter than expected. (b) Ramsey fringes for the two qubits. The extracted decoherence times $T_2$ are 8.7$\pm$0.3 $\mu$s and 4.6$\pm$0.2 $\mu$s for qubit 1 and qubit 2, respectively.    }
\end{figure}

The $T_1$ time for the qubits was measured by applying a $\pi$-pulse and measuring the state of the qubit as a function of delay time between the $\pi$- pulse and the readout pulse. Data are shown in FIG. \ref{Time_fig} (a). Relaxation times of 11.7 $\pm$ 0.2 $\mu$s and 2.1 $\pm$ 0.1 $\mu$s for q1 and q2 respectively were extracted.
The $T_1$ time of q2 is in good agreement with the expected Purcell limit. For q1, the expected Purcell limit is 20 $\mu$s. This is a factor of two greater than the observed $T_1$ time measured for this qubit.

The $T_2$ times were obtained from spin echo measurements and found to be 8.7 $\pm$ 0.3 $\mu$s and 4.6 $\pm$ 0.2 $\mu$s for q1 and q2 respectively (FIG. \ref{Time_fig}(b)). The $T_2$ time for q2 is found to be limited by its $T_1$ time. For q1 the $T_2$ is not limited by $T_1$, indicating that we have extra sources of decoherence affecting the qubits. 

The large capacitor pad architecture of the qubit gives it a large dipole moment, allowing for strong coupling to the microstrip. However, this also increases the radiation of the bare qubit. In the 3D architecture, the radiation loss is suppressed by the cavity resonator. A more compact way to suppress this loss is by placing the qubit in close proximity ({\it i.e.},  a small fraction of the radiation wavelength) to a conductive plane, as illustrated in FIG. \ref{Chip_fig}(b). The conducting plane generates a mirror image of the qubit dipole that radiates $\sim$180 degrees out of phase with the qubit dipole. The fields generated by the qubit and the image act to cancel each other, thereby suppressing the radiated power.

\begin{figure}[tb]
\includegraphics[width=8.6cm]{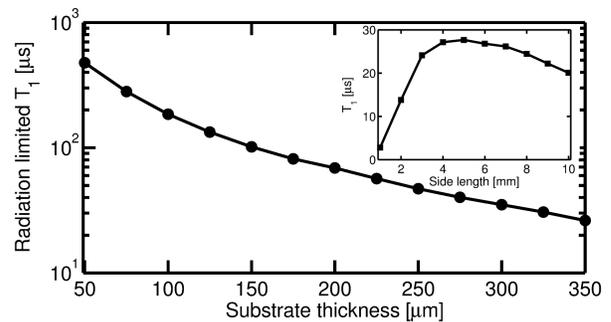}
\caption{\label{Rad_T1_fig} Calculated relaxation time due to radiation losses for a qubit above a superconducting plane on the back side of the substrate. The spacing between the qubit and the plane is set by the substrate thickness. The inset shows the expected relaxation time as a function of chip side length for a 350 $\mu$m thick substrate.}
\end{figure}

To quantify the effect of the conducting plane we use a finite element solver to calculate the time-averaged power flowing outwards from the dipole both with and without a superconducting plane. We find that the radiated power, $\overline{P}$, is suppressed by a factor of $\sim$100-400 in the far-field region for a substrate thickness of 350 $\mu$m over a 4-8 GHz frequency range. We estimate the effect of radiation loss on the relaxation time by defining a radiation resistance $R_{rad}=V^2/2\overline{P}$, where $V$ is the port voltage driving the dipole. The radiation limited relaxation time $T_{rad}=R_{rad}C_s$ is then calculated, where $C_s$ is the qubit shunt capacitance. Both the chip size and the substrate thickness affect the calculated $T_1$. The chip size dependence is shown in the inset of FIG. \ref{Rad_T1_fig} where the radiation limit on $T_1$ for a substrate thickness of 350 $\mu$m is shown. For a chip size of 5$\times$5 mm$^2$ we find $T_{rad}=26$ $\mu$s, two orders of magnitude higher than the expected $T_{rad}= 0.13$ $\mu$s for a device without the conductive plane. Moreover, as shown in the main panel of Fig. \ref{Rad_T1_fig}. there is a significant increase in the calculated radiation-limit on $T_1$ as the substrate thickness decreases. 

In this calculation, the effect of the sample box was not included. In order to evaluate the importance of the sample box for a qubit in close proximity to a superconducting ground plane, we measured the device with the lid removed from the box, thereby leaving the qubit and microstrip resonator exposed to the environment of the magnetic shielded, 20 mK stage of the dilution refrigerator. In these measurements we found $T_1$ and $T_2$ times of 9.7$\pm$0.5  and 8$\pm$0.5 $\mu$s, respectively. This shows that the sample box has minimal effect on the system.

Another important effect to consider is loss due to the materials. More specifically, potential sources of microwave loss at low powers and low temperatures are two level systems (TLSs) that are located primarily at the substrate-metal and substrate-vacuum interfaces \cite{Gao2008, Martinis2005, Wenner2011,Sandberg2012}. To quantitatively account for this, we calculated the filling factors of the different interfaces for qubit geometry and compared it to narrow gap CPW resonators previously studied \cite{Sandberg2012}. We found a decrease of $\approx$ 5 and $\approx$ 10 in the filling factors of the Si-vacuum and TiN-Si interfaces respectively for the qubit geometry. From the extracted filling factors we calculate the TLS limit of the $T_1$ time to be between 35 $\mu$s up to 300 $\mu$s, depending on the loss distribution between the two interfaces. This is well above our measured $T_1$ time, hence we do not believe that the TLSs are is a limiting factor. A final materials concern is the use of the Al-TiN hybrid circuitry. We note that earlier work has shown that there is no measurable increase in the loss of lumped element TiN resonators at low temperatures due to the TiN-Al interface\cite{Vissers2012}. In fact, that work showed that there is a significant reduction of quasiparticle loss due to the higher $T_C$ of the TiN relative to Al (~4.5 vs. 1.1 K). These data, therefore, indicate that our qubit lifetimes are not yet limited by the materials.

In conclusion our measured $T_1$ time is in agreement with the expected $T_1$ from the combined contributions of the Purcell effect due to the readout resonator (20 $\mu$s) and the calculated radiation limit (26 $\mu$s). The measurement shows that the backside superconducting plane acts to suppress the radiation loss, and the simulations predict that decreasing the distance between the qubit and the plane can significantly increase $T_1$. Based on these measurements and simulations, we believe that 2D, planar cQED circuits with lifetimes comparable to those of their 3D counterparts can be achieved.  In addition, having a local radiation suppression mechanism for the qubit makes it possible to bring in control lines to the qubit without reducing the qubit lifetime. This is a significant advantage compared to using 3D cavity resonators.\\

The authors are grateful for the input on the manuscript by Dan Slichter at NIST and the input by G\"{o}ran Joansson and Anton Frisk Kockum at Chalmers. We are also very grateful for the help of Mary Beth Rothwell at IBM T. J. Watson Research Center for assistance on the junction deposition.      
This work was supported by DARPA and the NIST Quantum Information initiative. This work is property of the US Government and not subject to copyright. 

\bibliography{BibTiNQubit}

\end{document}